\documentstyle[aps]{revtex}

\begin{document}
\draft
\title{About the influence of square-root Van Hove singularity on the
critical temperature of high-Tc superconductors}
\author{E. A. Pashitskii $^1$, V. I. Pentegov $^1$ and E. Abraham $^2$}
\address{$^1$ Institute of Physics, National Academy of Sciences of
Ukraine\\ Kiev, 252650, Ukraine}
\address{$^2$ Department of Physics, Heriot Watt University,\\
Edinburgh EH14 4AS, United Kingdom}

\date{To be published in JETP Letters v.67, No.7}
\maketitle

\begin{abstract}
It is shown that the square-root Van Hove singularity in the density of
states $\nu \left( E_F\right) \sim \left( E_F-E_0\right) ^{-1/2}$,
associated with the extended saddle-point features in the electronic
spectra of cuprate metal-oxides with hole-type conductivity, leads to
a nonmonotonic dependence of the critical temperature $T_c$ on the 
position of the Fermi level $E_F$ with respect to the bottom $E_0$ of 
the saddle. As the result of cancellation of the divergency of 
$\nu \left( E_F\right) $ in the electron-electron coupling constant, 
renormalized due to the account of strong coupling effects, $T_c$ 
approaches zero in the limit $E_F\rightarrow E_0$, contrary
to the case of the weak coupling approximation, which gives a finite (and
close to maximal) value of $T_c$ for $E_F\rightarrow E_0$. The dependence
of $T_c$ on the concentration of doped holes, obtained in the strong
coupling approximation, agrees qualitatively with experimental data for
the overdoped cuprate metal-oxides.
\end{abstract}

\pacs{PACS numbers: 74.20.-z, 74.25.Dw, 74.62.-c, 74.62.Dh.}

\noindent \makebox[\parindent][l]{\bf 1.}
The photo-emission spectroscopy experiments with high angular and energy
resolution \cite{Dessau,Abrikosov1,King,Gofron,Ding} show the existence of
the extended saddle point features (ESPF) near the Fermi level in the
electron spectra of the layered crystals of the cuprate metal oxides with
hole-type conductivity $\left( YBaCuO,\text{ }BiSrCaCuO,\text{ }
TlBaCaCuO\right) $. These so called ``flat'' portions of spectra should
give rise to a square-root Van Hove singularity in the electronic density
of states $\nu \left( E_F\right) \sim \left( E_F-E_0\right) ^{-1/2}$. The
Fermi energy $\mu _1\equiv \left( E_F-E_0\right) $ and the Fermi momentum
$ k_F\approx \sqrt{2m_1^{*}\mu _1}$ on these quasi one-dimensional
portions with effective mass $m_1^{*}\gg m_0$ (where $m_0$ is the free
electron mass) are anomalously small and for the optimally doped crystals
equal $\mu _1\simeq \left( 20\div 30\right) $ meV and $k_{F1}\simeq \left(
0.15\div 0.17\right) $ \AA $^{-1}$, while the extension of the saddle
anomalies in the momentum space is about $P_1\simeq 0.5$ \AA $^{-1}$
\cite{Gofron,Ding}.

As was shown in \cite{Abrikosov1,Abrikosov2} in the framework of the BCS
model \cite{Schrieffer}, the ESPF with square-root Van Hove singularities
lead, under the condition $\mu _1\leq T_c$, to the nonexponential
(power-like) $T_c$ dependence on the dimensionless coupling constant of
the effective interelectronic attraction $\lambda $: $T_c\approx \mu
_1\lambda ^2 $, where $\lambda \approx \lambda _1\sqrt{E_1/\mu _1}$, i.e.
the critical temperature approaches constant limit $T_c\approx E_1\lambda
_1^2$ when $\mu _1\rightarrow 0$. Thus, according to \cite{Abrikosov1},
for $E_1\geq 1$~eV it is possible to obtain rather high values of $T_c\geq
100$~K even with small constant $\lambda _1\sim 0.1$, regardless of the
underlying mechanism for Cooper pairing of the current carriers.

Yet for large values of $\lambda $ it is necessary to use the strong
coupling approximation \cite{Eliashberg}, which implies the
renormalization of the coupling constant $\widetilde{\lambda }=\lambda
/\left( 1+\lambda \right) $ leading to the cancellation of the divergency
$\lambda \sim \nu \left( \mu _1\right) \sim \mu _1^{-1/2}$ in the point
$\mu _1=0$. As shown in the present paper, due to this cancellation
$T_c\sim \mu _1$ when $\mu _1\rightarrow 0$, that is $T_c$ vanishes when
the Fermi level touches the bottom of the extended saddle anomalies,
contrary to the result of the weak coupling approach
\cite{Abrikosov1,Abrikosov2}, when $T_c$ has finite (and close to maximal)
value at $\mu _1=0$. The nonmonotonic dependence of $T_c$ on the
concentration of doped holes $n_p$ obtained with the account of the strong
coupling effects qualitatively agrees with the experimental data for the
overdoped cuprates \cite{Torrance,Tallon}.\\

\noindent \makebox[\parindent][l]{\bf 2.}
We shall proceed from the assumption that the high-temperature
superconductivity mechanism in the layered cuprate metal oxides is
governed by the Cooper pairing of the fermions (electrons, holes) through
the exchange of the virtual bosons (phonons, magnons, plasmons, excitons).
In the strong coupling approximation the superconducting state is
described by the set of equations for the normal $\Sigma _1$ and anomalous
$\Sigma _2$ self-energy parts. Under the condition that the characteristic
energy $ \tilde \Omega $ of bosons, mediating in the interaction, is much
greater then $T_c$, and with account for the quasi-2D character of the
electron spectrum in the layered cuprates and the anisotropy of the
electron-electron interaction, the linearized equation for the gap $\Delta
\left( {\bf k} ,\omega \right) $ on the Fermi surface for $T\rightarrow
T_c$ can be written as
\begin{equation}
\left( 1+\lambda \left( \theta \right) \right) \cdot \Delta \left( \theta
,0\right) =\frac 12\int\limits_0^{2\pi }\frac{d\theta ^{\prime }}{2\pi }
\int\limits_{-\tilde \Omega }^{\tilde \Omega }\frac{d\omega }\omega
\mathop{\rm Re} \Delta \left( \theta ^{\prime },\omega \right) \,\nu
\left( \theta ^{\prime },\omega \right) \,W\left( \theta ,\theta ^{\prime
},\omega \right) \tanh \frac \omega {2T_c},  \label{one}
\end{equation}
where $\lambda \left( \theta \right) =-\partial \Sigma _1\left( \theta
,\omega \right) /\partial \omega |_{\omega =0}$ is the dimensionless
constant of the retarded interaction, $\Delta \left( \theta ,\omega
\right) =\Sigma _2\left( \theta ,\omega \right) /\left( 1+\lambda \left(
\theta \right) \right) $ is the anisotropic gap in the quasiparticle
spectrum, $ \theta $ and $\theta ^{\prime }$ are the angles between one of
the main crystallographic axes (${\bf a}$ or ${\bf b}$) in the plane of
layers and the electron momenta ${\bf k}$ and ${\bf k}^{\prime }$ lying on
the Fermi surface, while $W\left( \theta ,\theta ^{\prime },\omega \right)
$ and $\nu \left( \theta ,\omega \right) $ are the matrix element of the
interelectron interaction and the electron density of states,
respectively, which could be expanded into the Fourier series in the
angles $\theta $ and $\theta ^{\prime }$. For the sake of simplicity we
shall use only the first terms in the expansion of $W\left( \theta ,\theta
^{\prime },\omega \right) $, corresponding to the representations A$_1$
and B$_1$ of the symmetry group C$ _{4v}$ of the cuprate plane $CuO_2$:
\begin{equation}
W\left( \theta ,\theta ^{\prime },\omega \right) =W_0\left( \omega \right)
+W_2\left( \omega \right) \cos 2\theta \cos 2\theta ^{\prime }+W_4\left(
\omega \right) \cos 4\theta \cos 4\theta ^{\prime },  \label{two}
\end{equation}
and we shall write the anisotropic density of states in the vicinity of
the closed and roughly cylindric Fermi surface (Fig.~1) as
\begin{equation}
\nu \left( \theta ,\omega \right) =\nu _{+}\left( \omega \right) +\nu
_{-}\left( \omega \right) \cos 4\theta ;\,\,\,\,\,\,\,\,\,\,\nu _{\pm
}\left( \omega \right) =\frac 12\left[ \nu _1(\mu _1)
\mathop{\rm Re}
\sqrt{\frac{\mu _1}{\mu _{1+}\omega }}\pm \nu _2\right] .  \label{three}
\end{equation}
Here $\nu _1(\mu _1)\sim \mu _1^{-1/2}$ is the density of states on the
quasi-1D portions of the Fermi surface near the extended saddle points
with square-root Van Hove singularity, and $\nu _2$ is the constant
density of states on the quasi-2D portions of the Fermi surface in the
direction of the diagonals of the Brillouin zone. The anisotropic coupling
constant $\lambda \left( \theta \right) $ is given in this case by
\begin{equation}
\lambda \left( \theta \right) =\lambda _0+\lambda _4\cos 4\theta
;\,\,\,\,\,\,\,\,\,\lambda _0=\nu _{+}\left( \omega \right) \cdot
W_0\left( 0\right) ;\,\,\,\,\,\,\,\,\,\lambda _4=\nu _{-}\left( \omega
\right) \cdot W_4\left( 0\right) .  \label{four}
\end{equation}

For sufficiently large positive values of $W_0\left( 0\right) $ and $
W_4\left( 0\right) $ the $s$-wave symmetry of the Cooper pairing may
prevail, resulting in superconducting state with the anisotropic gap
\begin{equation}
\Delta _s\left( \theta ,\omega \right) =\Delta _0\left( \omega \right)
+\Delta _4\left( \omega \right) \cos 4\theta .  \label{five}
\end{equation}
The critical temperature $T_c^s$ is given in this case by the solution of
the fallowing set of the coupled integral equations (to simplify it we
have neglected the frequency dependences of $\Delta $ and $W$ in the
frequency range $\left| \omega \right| <\tilde \Omega $):
\begin{eqnarray}
\left( 1+\lambda _0\right) \Delta _0+\frac 12\lambda _4\Delta _4 &=&\frac
12 W_0\left( 0\right) \int\limits_{-\tilde \Omega }^{\tilde \Omega }\frac{
d\omega }\omega \left[ \nu _{+}\left( \omega \right) \cdot \Delta _0+\frac
12 \nu _{-}\left( \omega \right) \cdot \Delta _4\right] \tanh \frac \omega
{ 2T_c^s};  \label{six} \\
\left( 1+\lambda _0\right) \Delta _4+\lambda _4\Delta _0 &=&-\frac 14
W_4\left( 0\right) \int\limits_{-\tilde \Omega }^{\tilde \Omega }\frac{
d\omega }\omega \left[ \nu _{+}\left( \omega \right) \cdot \Delta _4+\frac
12 \nu _{-}\left( \omega \right) \cdot \Delta _0\right] \tanh \frac \omega
{ 2T_c^s}.  \label{seven}
\end{eqnarray}

On the other hand, for a high enough positive value of $W_2$ the $
d_{x^2-y^2} $-wave symmetry of the Cooper pairing will dominate, with the
resulting gap given by $\Delta _d\left( \theta \right) \sim \cos 2\theta $
and the critical temperature determined by the equation
\begin{equation}
1+\lambda _0=\frac 14W_2\left( 0\right) \int\limits_{-\tilde \Omega }^{
\tilde \Omega }\frac{d\omega }\omega \left[ \nu _{+}\left( \omega \right)+
\frac 12\nu _{-}\left( \omega \right) \right] \tanh \frac \omega {2T_c^d}.
\label{eight}
\end{equation}
\\

\noindent \makebox[\parindent][l]{\bf 3.}
As the majority of experiments (see i.e.
\cite{Wollman,Tsuei,Kirtley,Norman} ) indicate the $d_{x^2-y^2}$-wave gap
symmetry in high-temperature superconductors, we shall consider only this
case, corresponding in our model to the high value of $W_2$ in
(\ref{two}). Taking into account (\ref {three}), the equation
(\ref{eight}) for $\mu _1<\tilde \Omega $ may be rewritten than as
\begin{equation}
1=\frac 14\frac{W_2}{1+\lambda _0}\cdot \left[ \frac 32\nu _1\left( \mu
_1\right) \int\limits_{-\mu _1}^{\tilde \Omega }\frac{d\omega }\omega
\sqrt{ \frac{\mu _1}{\omega +\mu _1}}\tanh \frac \omega {2T_c^d}+\nu _2\ln
\left( \frac{\tilde \Omega }{T_c^d}\right) \right] .  \label{nine}
\end{equation}
The first term in square brackets in (\ref{nine}) corresponds to the
quasi-1D portions of the Fermi surface with the square-root Van Hove
singularity, while the second term describes the influence of the quasi-2D
portions with the constant density of states.

The approximate integration over $\omega $ in (\ref{nine}) gives for $
T_c^d\ll \mu _1$,

\begin{equation}
\left( T_c^d\right) ^{3\alpha _1+\alpha _2}\approx \left( 4\mu _1\right)
^{3\alpha _1}\cdot \tilde \Omega ^{\alpha _2}\cdot \exp \left\{ -\frac{
1+\lambda _0}{\lambda _2}\right\} ,  \label{ten}
\end{equation}
where

\begin{equation}
\alpha _1=\frac{\nu _1\left( \mu _1\right) }{\nu _1\left( \mu _1\right)
+\nu _2};\quad \alpha _2=\frac{\nu _2}{\nu _1\left( \mu _1\right) +\nu
_2};\quad \lambda _2=\frac 12\left( \nu _1\left( \mu _1\right) +\nu
_2\right) \cdot W_2.  \label{eleven}
\end{equation}

In the opposite case of $T_c^d\gg \mu _1$, when $\nu _1\gg \nu _2$ and $
\lambda _0\approx \frac 12\nu _1W_0$ at $\mu _1\rightarrow 0$, the
cancellation of $\nu _1$ in the renormalized coupling constant reduces
(\ref {nine}) to

\begin{equation}
T_c^d\approx 2\mu _1\cdot \left( \frac{3W_2}{2W_0}\right) ^2,
\label{twelve}
\end{equation}
so that $T_c^d\rightarrow 0$ when $\mu _1\rightarrow 0$. It is possible to
show using (\ref{six}) and (\ref{seven}) that in the strong coupling
approximation the similar result ($T_c^d\sim \mu _1$ when $\mu
_1\rightarrow 0$) is valid for the $s$-wave gape symmetry as well,
contrary to the result of \cite{Abrikosov1,Abrikosov2}, obtained in the
weak coupling approximation, when $T_c$ is finite at $\mu _1=0$ and close
the maximal value.

The results of the numerical solution of eq. (\ref{eight}) are shown in
Fig.~2. The $T_c^d$ dependencies on $\mu _1$ are represented for several
values of parameters $\tilde \Omega $ and $\lambda _1=\nu _1^{*}W_2$,
where $ \nu _1^{*}\equiv \nu _1\left( \mu _1^{*}\right) $ and $\mu
_1^{*}\approx \left( 0.02\div 0.03\right) $~eV, which corresponds to the
position of the Fermi level in the optimally doped cuprates
\cite{King,Gofron}. Solid curves 1 and 2 are calculated for $\tilde \Omega
=0.1$~eV and $\tilde \Omega =2$~eV respectively, with the fixed ratio $\nu
_1^{*}/\nu _2=5$. The constants $ \lambda _1$ and $\lambda _0$ where
chosen so as to produce the position of the maximum of $T_c^d$ in the
point $\mu _1^{*}=0.03$~eV and the maximal value of $T_c\approx 110$~K,
experimentally observed in $BSCCO$ compounds.  Notice that for the fixed
ratio of the constants $\lambda _1$ and $\lambda _0\left( \mu
_1^{*}\right) $ these constraints on the position and value of the $T_c$
maximum lead to the power-law dependence of the constant $\lambda _1$ on
$\tilde \Omega $ with a small exponent $\beta \approx -0.06$, shown with
the log-log plot of the inset in Fig. 2. The value of the constant $
\lambda _1$, necessary for achieving sufficiently high $T_c$, is
relatively small and only weakly depends on the characteristic energy of
the interaction, determined by the specific mechanism of the Cooper
pairing.  With the dashed curves in Fig. 2 are shown the dependencies of
$T_c^d$ calculated for the same parameters, but without the
renormalization factor $ \left( 1+\lambda _0\right) $, which corresponds
to the weak coupling approximation \cite{Abrikosov1,Abrikosov2}.

Fig. 3 represents the concentrational dependencies of $T_c^d$,
corresponding to the curves 1 and 1$^{\prime }$ of Fig. 2, obtained with
account for the expression for the density of states (\ref{three}). The
theoretical curves are compared to the experimental values, taken from
\cite{Wollman}, of the critical temperature for different hole
concentrations $n_p$ per $Cu$ atom .  We see the good agreement of the
experimental data for the overdoped cuprates with the theoretical
dependence $T_c\left( n_p\right) $, obtained in the strong coupling
approximation.

In summary, we have shown that the effect of superconducting
critical temperature decreasing with the increase in the hole
concentration in the overdoped cuprates, with $T_c$ eventually becoming
zero at some concentration, is closely connected to the existence of the
square-root Van Hove singularity in the electronic density of states. This
conclusion is a consequence of the coupling constant renormalization due
to the strong coupling effects, and does not depend on the specific
mechanism of Cooper pairing and the symmetry of superconducting order
parameter.\\

The authors express their gratitude to A.~V.~Gurevich and A.~V.~Semenov
for useful discussions. This work was supported by grant 2.4/561 of the
State Foundation for the Fundamental Research of Ukraine and by grant
GR/L25363 (Visiting Fellowship Research Grant) of the EPSRC, United
Kingdom.

\begin{center}
{\bf Figure captions}
\end{center}

\begin{description}
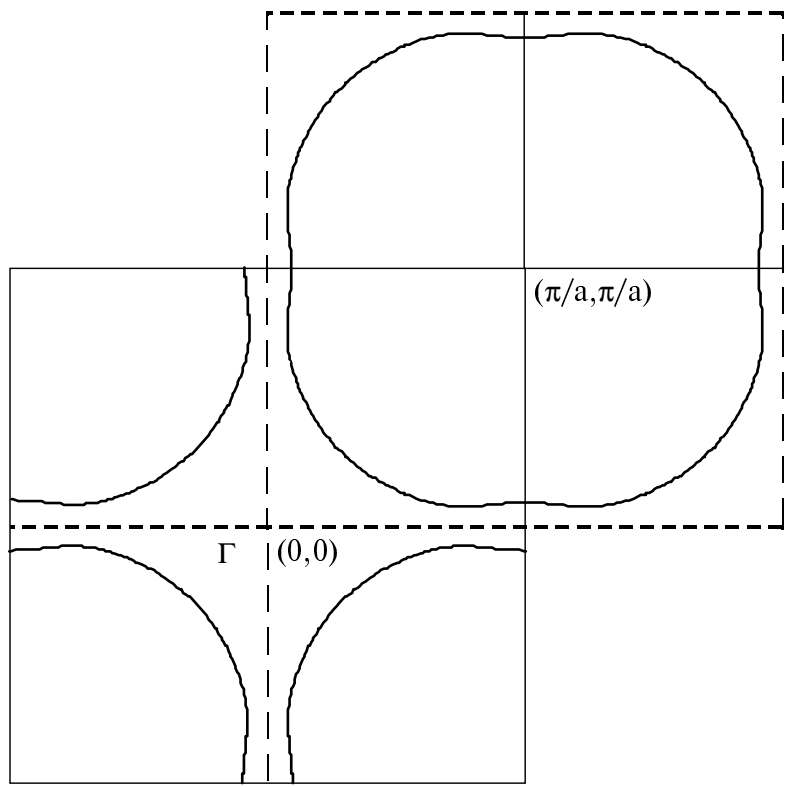
\item[Fig. 1]  Cross section of the Fermi surface of the cuprate layered
metal-oxides of the $BSCCO$ type. The closed hole-like Fermi surface is
centered in the corner $\left( \pi /a,\pi /a\right) $ of the Brillouin
zone.

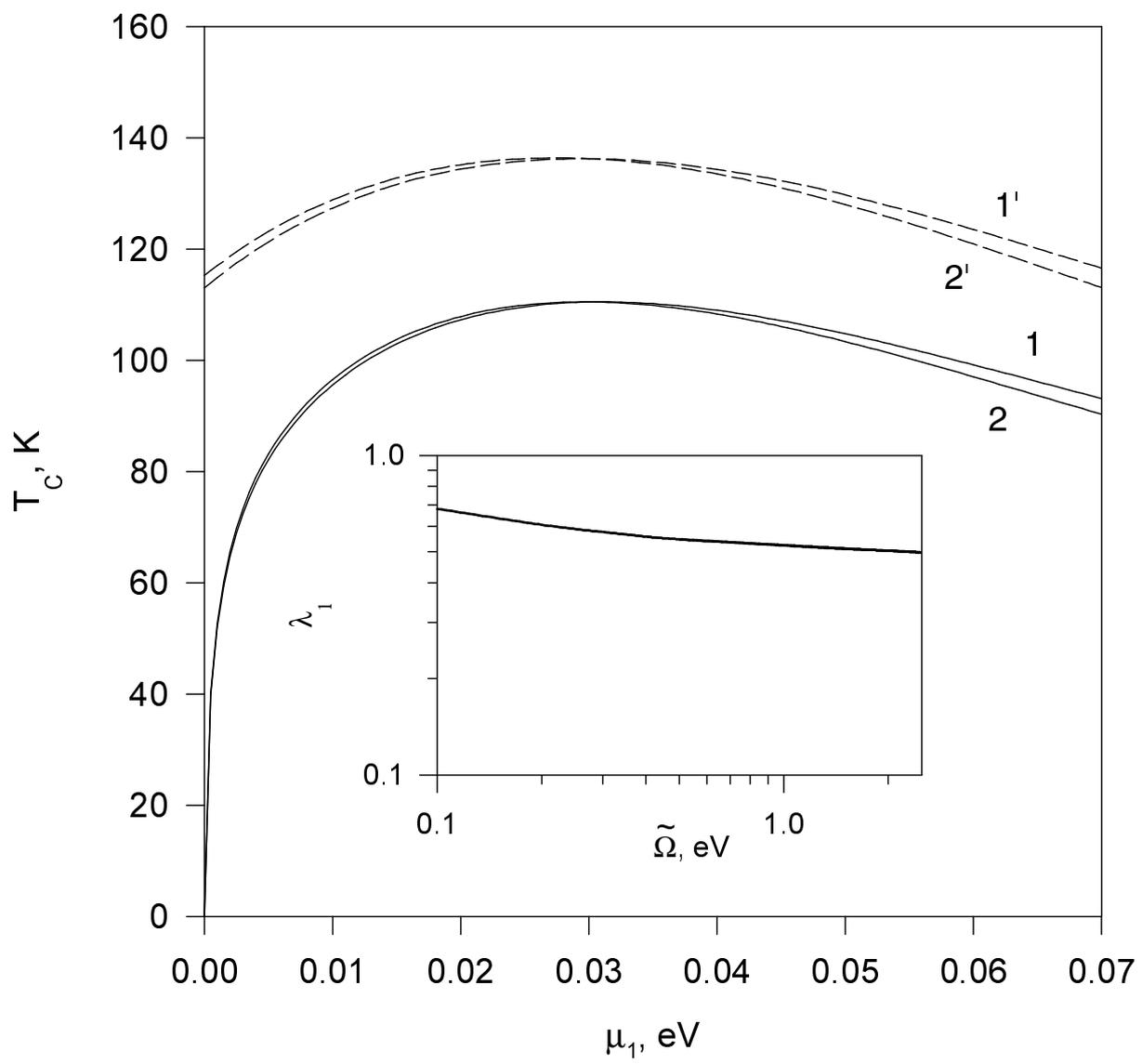
\item[Fig. 2]  Critical temperature $T_c$ in a function of the Fermi
energy $ \mu _1=E_F-E_0$ on the extended saddle point features, calculated
in the strong coupling (solid curves) and weak coupling (dashed curves)
approximations. Curves 1 and 1$^{\prime }$ correspond to $\tilde \Omega
=0.1$ ~eV and $\lambda _1=0.68$, curves 2 and 2$^{\prime }$ -- to $\tilde
\Omega =2 $~eV and $\lambda _1=0.5$.

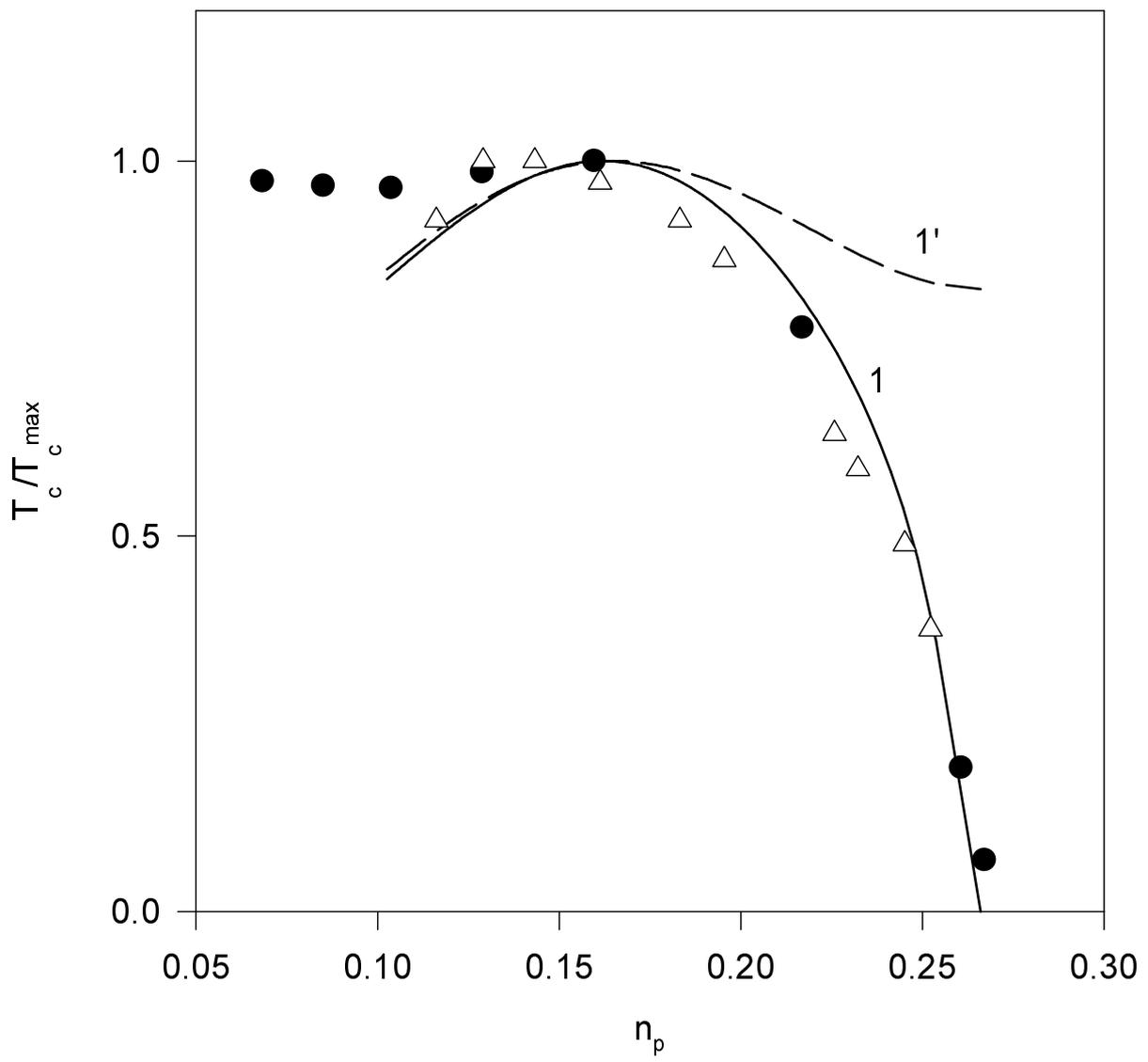
\item[Fig. 3]  Critical temperature $T_c$ in a function of the hole
concentration per $Cu$ atom $n_p$. Theoretical curves correspond to curves
1 and 1$^{\prime }$ of Fig. 2. The experimental data are taken from \cite
{Wollman}: ${\displaystyle }\bullet $ -- $TlPbCaYSrCuO\,\left( 1212\right)
$ , ${\scriptstyle \Delta }$~--$~BiPbSrLaCuO\,\left( 2201\right) $
\end{description}
\end{document}